\documentclass[12pt,preprint]{aastex}

\shorttitle{Wide Substellar Companion to an M-dwarf}
\shortauthors{Radigan et al.}

\newcommand{\msun}{\ensuremath{M_{\odot}}}

\begin{document}

\title{Discovery of a Wide Substellar Companion to a Nearby Low-Mass
  Star}

\author{Jacqueline Radigan, David Lafreni\`ere, Ray Jayawardhana}
\affil{Department of Astronomy and Astrophysics, University
  of Toronto, 50 St. George Street, Toronto, ON, M5S 3H4, Canada}
\author{Ren\'e Doyon}
\affil{D\'epartement de Physique and Observatoire du Mont M\'egantic, Universit\'e de Montr\'eal, C.P. 6128, Succ. Centre-Ville, Montr\'eal, QC, H3C 3J7, Canada}
\email{radigan@astro.utoronto.ca}

\begin{abstract}

We report the discovery of a wide ($135\pm25$~AU), unusually blue L5
companion 2MASS J17114559+4028578 to the
nearby M4.5 dwarf G 203-50 as a result of a targeted search for common
proper motion pairs in the Sloan Digital Sky Survey and the Two
Micron All Sky Survey.  Adaptive Optics imaging with Subaru indicates that neither
component is a nearly equal mass binary with separation $> 0.18\arcsec$, and
places limits on the existence of additional faint companions.  An examination of TiO and CaH features
in the primary's spectrum is consistent with solar metallicity and
provides no evidence that G 203-50 is metal poor.  We estimate an
age for the primary of 1-5 Gyr based on activity.  Assuming
coevality of the companion, its age, gravity and metallicity can be
constrained from properties of the primary, making it a suitable
benchmark object for the calibration of evolutionary models and for
determining the atmospheric properties of peculiar blue L dwarfs.
The low total mass ($M_{tot}=0.21\pm0.03$~\msun), intermediate mass
ratio ($q=0.45\pm0.14$), and wide separation of this system demonstrate that the star
formation process is capable of forming wide, weakly bound binary systems
with low mass and BD components.  Based on the sensitivity
of our search we find that no more than $2.2\%$ of early-to-mid M dwarfs ($9.0
<M_V < 13.0$) have wide substellar companions with $m>0.06$~\msun.
  
\end{abstract}

\keywords{binaries: general --- stars: formation --- stars: individual
  (2MASS J17114559+4028578, G 203-50) stars: low-mass, brown dwarfs}

\section{Introduction}\label{sect:intro}

Although star birth is a complex process, the
observation of binary systems--frequencies, mass ratios,
and separations--can provide insight into the formation process as well as
constraints for theoretical models. 
The formation of brown dwarfs (BDs) is particularly challenging since their masses are an order of
magnitude smaller than the typical Jeans mass in molecular clouds.

Whether BDs form similarly to their more massive stellar counterparts
or require additional mechanisms is currently an open question.  The answer may lie in the multiplicity
properties of these substellar objects. 
Whereas free-floating BDs are observed in abundance, finding BDs as companions to stars has
proved more difficult.  A ``brown dwarf desert'' ($\lesssim0.5\%$ companion fraction) is observed at
close separations ($<3$~AU) to main sequence stars,  in comparison to
a significant number of both planetary and stellar mass companions
seen at similar separations \citep{marcy00}.  It has recently been
determined that this desert does not extend out to larger separations for solar analogs (F,G,K stars),
$\sim7\%$ of which are found to harbor substellar companions at
separations greater than 30~AU \citep{met05}.
However, searches for substellar companions to M dwarfs at large
separations ($\gtrsim$ 40 AU) have yielded mostly null results
\citep[e.g.][]{allen08,mccarthy04,hinz02,daemgen07} or sparse results
\citep[e.g.][]{oppen01}.  There are only 5 known BD
companions to stellar M dwarf primaries at separations greater than 40AU: TWA 5b,c
\citep{lowrance99}, G 196-3B \citep{rebolo98}, GJ 1001B
\citep{goldman99}, Gl 229B \citep{nakajima95}, and LP 261-75B
\citep{burgasser05,reid06}.  For thoroughness we note that the L2.5 companion GJ 618.1B
\citep{wilson01} may also fall into this category, however it is
more likely stellar.  In the VLM regime
($M_{1}<0.1$~\msun) surveys have found that no more than $\sim1\%$ of
stars have wide companions, including stellar ones
\citep{burgasser07a}.  Additionally, VLM binaries are found to be on
average 10-20 times more tightly bound than their stellar counterparts, hinting
that disruptive dynamical interactions may play an important role in
their formation \citep{close03}. These observations have been cited as evidence in favor of
the ejection hypothesis \citep{rep01, bate05} where BDs and VLM stars
are thought to be stellar embryos formed by the fragmentation of a
more massive pre-stellar core, then prematurely ejected from their
birth environments.  However, BD companions to more massive stars do not
tend to form harder binaries than stellar systems of similar total
mass \citep[e.g.][]{reid01,met05}.  While this is potentially evidence
that BDs can form similarly to stars via turbulent fragmentation
within molecular clouds \citep{padoan02}, it is also consistent with simulations of
disk instabilities \citep[e.g.][]{stamatellos07,boss00}, which are capable of producing
substellar companions around more massive primaries.

While a significant fraction of solar mass stars may retain wide
BD companions, this does not seem to hold true for lower mass stars.
As a result, very few wide BD companions to low mass stars are known.
The discovery and characterization of these systems, especially in the intermediate
range between the solar analog and VLM regimes, will help complete the
emerging  picture of BD multiplicity at wide separations.  
Additionally, wide BD companions to stars make suitable
``benchmark'' objects, as their properties can be
inferred from those of the primary \citep{pinfield06}.  This is important for the
calibration of BD evolutionary models, which requires independent age estimates.

Here we present the discovery of a wide substellar companion to a nearby M4.5
star.  The search, discovery and followup observations are outlined in \S\ref{sect:search}, while the
physical properties of the system and its components are given in \S
\ref{sect:properties}.  In \S\ref{sect:discussion} we
discuss the companion's unusual NIR colors, possible formation
scenarios, and the sensitivity of our search.  Given a space
density for M dwarfs we make a crude estimate of how rare such systems
may be. A brief summary and outlook are presented in \S \ref{sect:conclusions}.

\section{Discovery and Observations}\label{sect:search}

The binary G~203-50 / 2MASS J17114559+4028578 (G~203-50AB hereafter) 
was found in a cross-match of the SDSS DR6 Photoprimary Catalogue
\citep{sdss} and
2MASS Point Source Catalogue \citep{skrutskie06} in which we searched for common proper
motion pairs containing at least one VLM or BD component.  The
cross-correlation of catalogues, calculation of proper motions, and the
identification of co-moving stars was done in parallel for 4~$\rm{deg^2}$
sections of the sky at a time, spanning the contiguous region of the
SDSS Legacy survey in the northern galactic cap.  We made preliminary cuts to include only 2MASS
sources with S/N $>$ 5 in at least one band (J, H, or K) and not
flagged as minor planets, and SDSS
sources that were not classified as ``sky pointings'' or electronic
ghosts.  For every 2MASS source the closest SDSS match was found and
proper motion vectors with uncertainties were computed.  A cut was made in order to select only
stars that had moved at the $3\sigma$ level compared to all other
stars within the area.
Stars within 120\arcsec~of one another with proper motion amplitudes
agreeing within 2$\sigma$ and proper motion components agreeing within
1$\sigma$ in one of RA and DEC, were flagged as potential binaries.
We applied a color cut of
$z^{\prime}-J >2.5$ for at least one component to the proper-motion-selected
sample in order to search for BD companions.  Finally all candidates
were examined visually to eliminate artifacts and spurious
matches.  Of all candidate systems, G~203-50AB stood out as harboring
a very red companion with $z^{\prime}-J\approx2.9$.
Although at first glance the companion passed our color
cuts, the quoted $z^{\prime}$ error of $\pm2$ magnitudes rendered this color
meaningless.
Fortunately the primary was of known spectral type and absolute magnitude
$M_J=9.34\pm0.18$ \citep{mtcn7}, yielding $M_J=13.3$ for the
secondary, assuming it to be at the same distance as the primary.  Average absolute magnitudes
as a function of spectral type \citep{dahn02} suggested that the
companion was indeed a mid-L dwarf.

\begin{figure}[here]
\epsscale{1}
\plotone{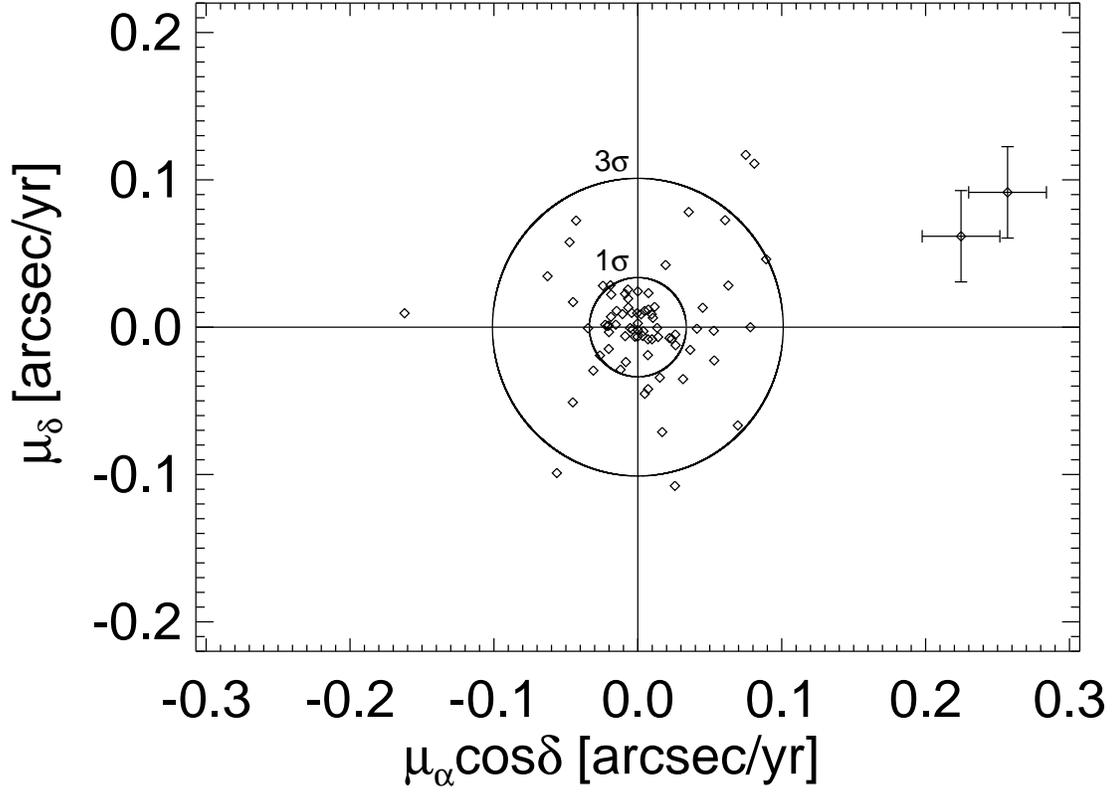}
\caption{Proper motions of all sources within $10\arcmin$ of G~203-50.  Circles indicate the $1\sigma$
  and $3\sigma$ dispersions.  G203-50 and 2M1711+4028 are displayed
  with error bars based on the $1\sigma$ dispersions along each axis,
  which dominates over the astrometric errors quoted by the individual
  catalogues.  2MASS and SDSS observations of this system are separated by a baseline of 6.0959 years.
\label{fig:pm}}
\end{figure}

As a verification of the system's physical association we have plotted
the proper motions of G~203-50AB along with all other stars
within 10\arcmin~of the primary.  Figure
\ref{fig:pm} clearly shows G~203-50AB as a co-moving pair.

In order to establish a spectral type for the companion we obtained an
infrared  spectrum (R$\sim$750) of 2MASS~J17114559+4028578
(2M1711+4028~hereafter) on 2008 Feb 28 using the
SpeX Medium-Resolution Spectrograph \citep{spex} at NASA's Infrared
Telescope Facility (IRTF). Observations were made in short wavelength
(0.8 to 2.5 $\mu$m) cross-dispersed mode with the $0.8\arcsec$ slit and the seeing was
0.7$\arcsec$-0.9$\arcsec$. We obtained eight 180~s exposures arranged in two ABBA nod
patterns with a nod step of 7$\arcsec$ along the slit. For telluric and
instrumental transmission correction, the A0 star HD165029 was
observed immediately after the target at the same airmass.
Flat-fielding, background subtraction, spectrum extraction,
wavelength calibration, order merging and telluric correction were
done using SpeXtool \citep{spextool1,spextool2}.  No scaling was
applied to the cross-dispersed spectrum when merging the orders.  The
spectrum is presented and analyzed in \S\ref{sect:secondary}.

To constrain the possible binary nature of the primary and the
companion, adaptive optics imaging observations of the system were
obtained on 2008 July 8 at the Subaru telescope (open use program
S08A-074). The observations were made with the AO36 adaptive optics
system \citep{takami03} and the CIAO near-infrared camera
\citep{murakawa04}. The primary star G~203-50 was used for wave front
sensing. Five exposures of 7~s $\times$ 3~co-additions were obtained
in $K_{\rm s}$ over five dither positions. The images were reduced in
a standard manner. A sky frame was obtained as the median of the five
images after masking the regions dominated by the signal from the
target. After subtraction of this sky frame, the images were divided
by a normalized dome flat image and bad pixels were replaced by a
median over their neighbors. All images were finally co-aligned,
flux-normalized, and co-added. Owing to the faint $R$-band magnitude
of the target star, the adaptive optics correction achieved is rather
poor, with a PSF FWHM of 0.18\arcsec.  Visual inspection of the adaptive optics images indicates that neither the primary nor the companion
is a nearly equal-mass binary with a separation of $\sim$0.18\arcsec\ or larger. Subtraction of a properly shifted and scaled version of the primary star PSF from that of the companion confirms that conclusion as this operation leaves no obvious residual.  Our adaptive
optics images also provide constraints on the existence of additional, fainter companions in the system. The detection limits achieved
indicate that the primary has no other companion with $\Delta K_{\rm
  s}<4.3$, 7.0 and 7.8 mag at separation greater than 0.5\arcsec,
1\arcsec\ and 2\arcsec respectively.  
Similarly, the secondary has no companion with $\Delta K_{\rm s}<4.3$ above 0.5\arcsec.

\section{Physical Properties}\label{sect:properties}
A summary of observational and physical properties of
the system is given in tables 1 and 2.  
Proper motions for each component, in a reference frame defined by the
median proper motion of all background stars within $10\arcmin$, were
found and averaged to give a mean system proper motion of
$242\pm15$~mas~${\rm yr^{-1}}$ in right ascension and
$77\pm17$~mas~${\rm yr^{-1}}$ in declination. This is in good agreement with the proper
motion tabulated for G203-50 of $250.5\pm5.5$~mas~${\rm yr^{-1}}$ and
$84.2\pm5.5$~mas~${\rm yr^{-1}}$ in the Revised NLTT catalogue \citep{nltt}.  

\subsection{The primary: G 203-50} \label{sect:primary}
\begin{figure}[here]
\epsscale{1}
\plotone{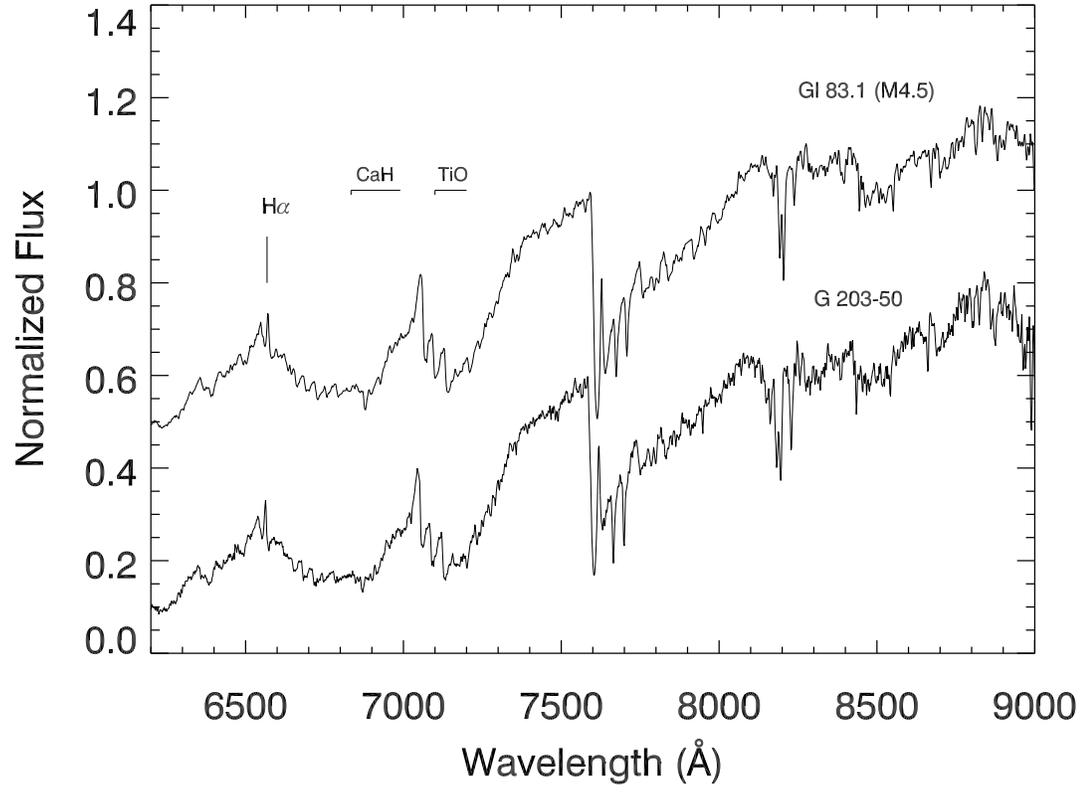}
\caption{Spectrum of G 203-50 \citep{mtcn7}, plotted alongside
  the M4.5 spectral standard Gl 83.1 for comparison.
\label{fig:mspec}}
\end{figure}

G~203-50 has been assigned a spectral type of M4.5 and an absolute
magnitude of $M_J=9.34\pm0.18$\footnote{errors were not provided with the
  online data, but we have inferred them from the quoted error in
  photometric distance} by \citet{mtcn7}.  The spectral type
was assigned based on the TiO5 spectral index \citep{cruz02}, while the
absolute J magnitude is derived from polynomial fits to the TiO5, CaOH, and CaH2
spectral indices as a function of absolute J magnitude for M dwarfs with trigonometric
parallaxes \citep{cruz02}.   We have verified the
spectral type by comparison of G 203-50's spectrum (courtesy of Neill
Reid) with reference spectra from \citet{kirkpatrick91} using standard
spectra made available online by Kelle Cruz\footnote{http://www.astro.caltech.edu/$\sim$kelle/M\_standards/}, and
\citet{pickles98}. The spectrum of G 203-50 is shown alongside the M4.5
spectral standard Gl 83.1 in figure \ref{fig:mspec}.
  
Since the absolute J magnitude is derived from spectral indices which
may vary continuously between spectral types, it is in principle more
precise than a magnitude derived based on the average value for all
members of a given
subtype. However, the absolute J magnitude provided by \citet{mtcn7} seems to have
unrealistically low errors, given the large spread in absolute magnitude
for mid-M spectral types \citep[e.g. see figure 4 of ][]{cruz02}.  As
a sanity check we conducted a SIMBAD search yielding 45 M4.5 dwarfs with
measured parallaxes and J band fluxes.  The distribution of absolute J
magnitudes has a median value of 9.0, and a standard deviation of
0.44.  Therefore we choose to adopt a more conservative error estimate of 0.44 magnitudes.

Using $M_J=9.34\pm0.44$, the corresponding photometric distance is
$22.2\pm4.5$~pc.  Assuming an age $> 500$~Myr we derived a mass of $0.146\pm0.031$~\msun ~using the empirical mass-luminosity relationship of
\citet{delfosse2000}, also in agreement with the models of
\citet{baraffe98}.

The spectrum of G 203-50 shows moderate $H\alpha$ emission.  We
measured the $H\alpha$ equivalent width (EW) to be
$3.8\pm0.5$~\AA~ with errors stemming from a high sensitivity to the
regions chosen to fit the pseudo-continuum on either side of the
line. We used a cubic polynomial to fit the pseudo-continuum regions
from  $6544.3$-$6551.9$~\AA~and from $6576.6$-$6582.3$~\AA.
The relationship between $H\alpha$ activity and age becomes degenerate
for low mass dwarfs \citep{zuckerman04} making it difficult to
draw firm conclusions about the age of G203-50.  However,
the activity lifetime of M stars in terms of $H\alpha$ emission has
recently been constrained using a sample of 38000 M dwarfs from SDSS
Data Release 5 \citep{west08}.  The activity lifetime for M4 and
M5 stars respectively is found to be 4.5 and 7~Gyr.  Considering that
G 203-50 is more active than the majority of M dwarfs of similar
spectral type \citep[e.g.][see online data]{mtcn7} it is reasonable to
infer that it is not near the end of its active phase.  Nor does
G 203-50 display signs of extreme youth, as it has no associated
x-ray source in ROSAT.  Therefore we tentatively estimate the age of
G 203-50 to be between 1 and 5 Gyr.

Based on the metallicity scale of \citet{gizis97} the TiO5 and CaH2
indices of G 203-50 are consistent with solar metallicity, indicating
[M/H]$>$ -1.0.  We measured the metallicity index of \citet{lepine07} to
be $\zeta_{{\rm TiO/CaH}}=0.95$, where solar metallicity is represented by
$\zeta_{{\rm TiO/CaH}}=1$, and metal poor stars have $\zeta_{{\rm
    TiO/CaH}}<0.825$.  Our own fits of model spectra in the 6200-7300~\AA~
region verify these results.  We fit the NextGen99 model atmospheres
\citep{hauschildt99} to G 203-50's spectrum in the region of TiO and
CaH bands of 6200-7300~\AA.  We used a grid of models with $3000~K <
T_{eff} < 3300 K$ in steps of 100K, $4.5 < \log g < 5.5$ in steps of 0.5 dex, and $-2.0 < [M/H] < 0.0$
in steps of 0.5 dex.  These parameter ranges were chosen based on
values of $T_{eff}=3114\pm125$ and $\log g =5.14\pm0.05$ computed from the evolutionary models of
\citet{baraffe98} using the 1-5 Gyr isochrones.  In each case the templates were convolved with a
gaussian with a FWHM equal to the instrumental resolution ($\sim5.5$~\AA) of G 203-50's spectrum
and then interpolated onto the data.  As noted by \citet{lepine07}, none of the model spectra
were good fits, with the CaH band consistently appearing too strong
relative to TiO.  However, as expected the best fitting model spectra were those with solar
metallicity, with the fit becoming progressively worse for decreasing
[M/H].  Therefore we find no evidence to suggest that G 203-50 is
metal poor.  A higher resolution spectrum is required to determine the
metallicity more precisely \citep[for example, using the method of][accurate to $\sim0.1$~dex]{bean06}.

\subsection{The Companion: 2MASS J17114559+4028578} \label{sect:secondary}
\begin{figure}[here]
\epsscale{1}
\plotone{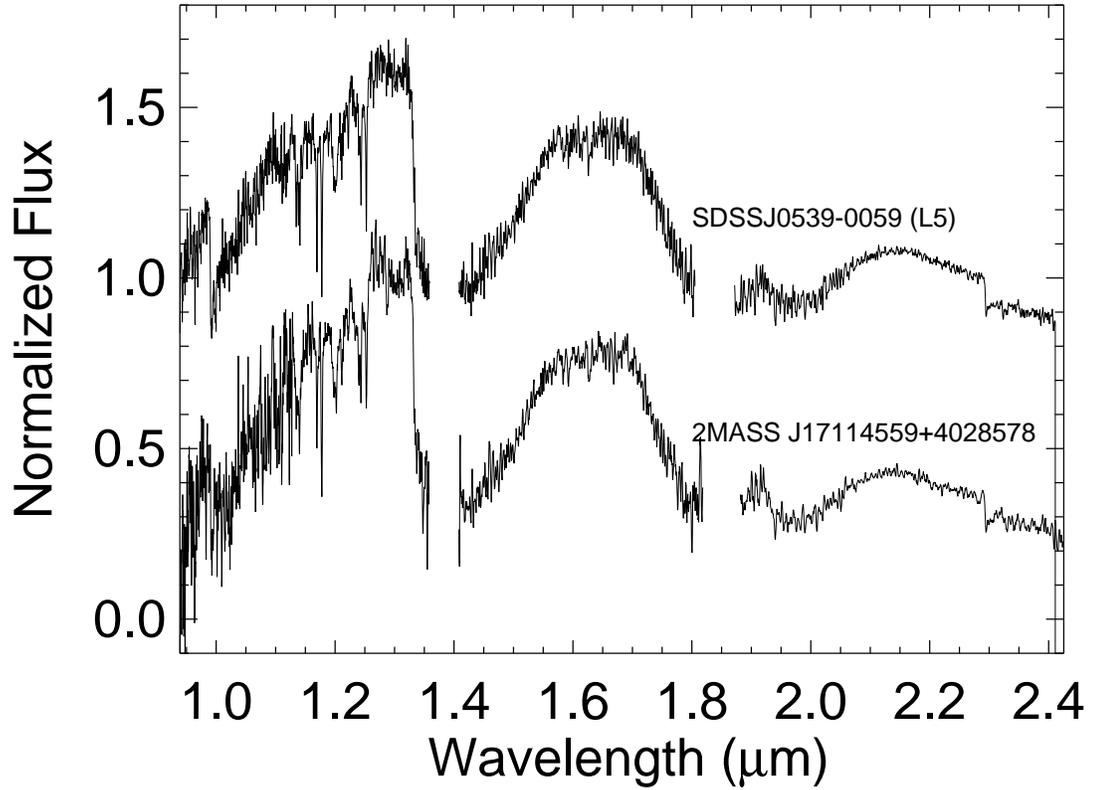}
\caption{Spectrum of 2M1711+4028 obtained at IRTF using the SpeX
  Medium-Resolution Spectrograph.  The spectrum is plotted alongside
  the L5 dwarf SDSSp~J05395199-0059020 \citep{fan00}, from the IRTF Spectral
  Library.  Both spectra are normalized in 1.27-1.33~$\mu$m and offset for clarity.
\label{fig:spectrum}}
\end{figure}

Using reference spectra from the IRTF\footnote{http://irtfweb.ifa.hawaii.edu/$\sim$spex/spexlibrary/IRTFlibrary.htm} (maintained by John Rayner), SpeX
Prism (maintained by Adam Burgasser), CGS4 \citep{leggett2001}, and NIRSPEC \citep{mclean03}
spectral libraries, we determined the best fitting spectral type for
the companion to be L5 (see figure \ref{fig:spectrum}).   However, other reasonable
matches were found from L3.5 to L6.5.  We also measured spectral
indices defined by \citet{geballe02}, \citet{tokunaga99},
\citet{reid01b}, and \citet{mclean03}.  The indices are presented in table \ref{tbl:tab3}, yielding
spectral types from L4.5 to L8.  We calculated synthetic 2MASS colors of $J-K_s$=1.28, $J-H$=0.79 and $H-K_s$=0.49
from the spectrum\footnote{the quoted
2MASS $J$, $H$, and $K_s$ magnitudes are
flagged as being biased by the nearby primary, however our derived
colors agree with the former to within 0.1 mag} of 2M1711+4028 using the relative
spectral response curves and zero-magnitude fluxes given in
\citet{cohen03}.  These NIR colors make 2M1711+4028 unusually blue for
a mid-late L dwarf \citep{cushing08}.
The large range in spectral type implied by the spectral indices, and
the unusually blue NIR colors of 2M1711+4028 likely have a common origin.
Several instances of anomalously blue L dwarfs have been previously 
noted \citep[e.g.][]{knapp04,burgasser08}, where the
optical spectra appear normal, but the NIR colors are much bluer than
average.  These blue L dwarfs are characterized by enhanced H2O absorption
bands and diminished CO, giving the appearance of a later spectral type in the NIR,
while J and K band features such as FeH are more consistent with an
earlier optical classification (see \S \ref{sect:blue} for further discussion).  Similarly for 2M1711+4028, the strong H2O indices predict
spectral types from L6.5 to L8 while z-FeH and J-FeH indices predict types
$<$L6 and the K1 (not to be confused with \ion{K}{1}) index predicts a spectral type of L4.5.  This is consistent with the two best matching
reference spectra: the relatively blue L dwarfs
SDSSp J05395199-0059020 \citep{fan00} and 2MASS~J15074769-1627386 \citep{reid00}, both of which
have optical spectral types of L5, and the latter of which is an optical spectral
standard.  Considering all of our measurements, we have
assigned a NIR spectral type of $L5^{+2}_{-1.5}$ to 2M1711+4028, with our choice being most
strongly influenced by the best fitting reference spectra.   
The error bars span the entire range of reasonable spectral types based on
template fitting and spectral indices, excluding the H2O indices.

Using the absolute magnitude-spectral type relationship provided by
\citet{dahn02} an absolute magnitude of $13.5^{+0.7}_{-0.5}$ is found for
the secondary, corresponding to a distance of $20.0\pm6.3$~pc.  This
is consistent with the distance of $22.2\pm4.5$~pc derived for the
primary. The average distance for the system is $21.2\pm3.9$~pc.   
We find an effective temperature of $1700^{+210}_{-250}$~K based on the
spectral type-effective temperature relationship of
\citet{golimowski04}.  Using the DUSTY model isochrones for 1 and 5~Gyr
\citep{chabrier00}, we derive a mass of
$0.066_{-0.015}^{+0.008}$~\msun.  The quoted uncertainty takes into account the 1$\sigma$ uncertainty
in $ T_{eff}$ as well as the age interval.  While our upper limit of
$0.074$~\msun~straddles the stellar-substellar boundary, this is the most
conservative estimate, allowing for a very broad range in spectral
types.  We conclude that 2M1711+4028 is most likely substellar, an
issue which can be resolved in the future by obtaining an
optical spectrum in order to further constrain its spectral type.
 
Based on 2MASS and SDSS astrometry 2M1711+4028 is separated from
G~203-50 by a mean angular separation of $6.47\arcsec\pm0.14\arcsec$, in agreement with the angular separation measured from the
adaptive optics images of $6.40\arcsec \pm 0.02\arcsec$ at $234.1\degr
\pm 0.2\degr$.  This corresponds to a projected separation of $135\pm25$~AU, at the average
system distance.

\section{Discussion}\label{sect:discussion}

\subsection{The blue NIR colors of 2M1711+4028}\label{sect:blue}
\begin{figure}[here]
\epsscale{1}
\plotone{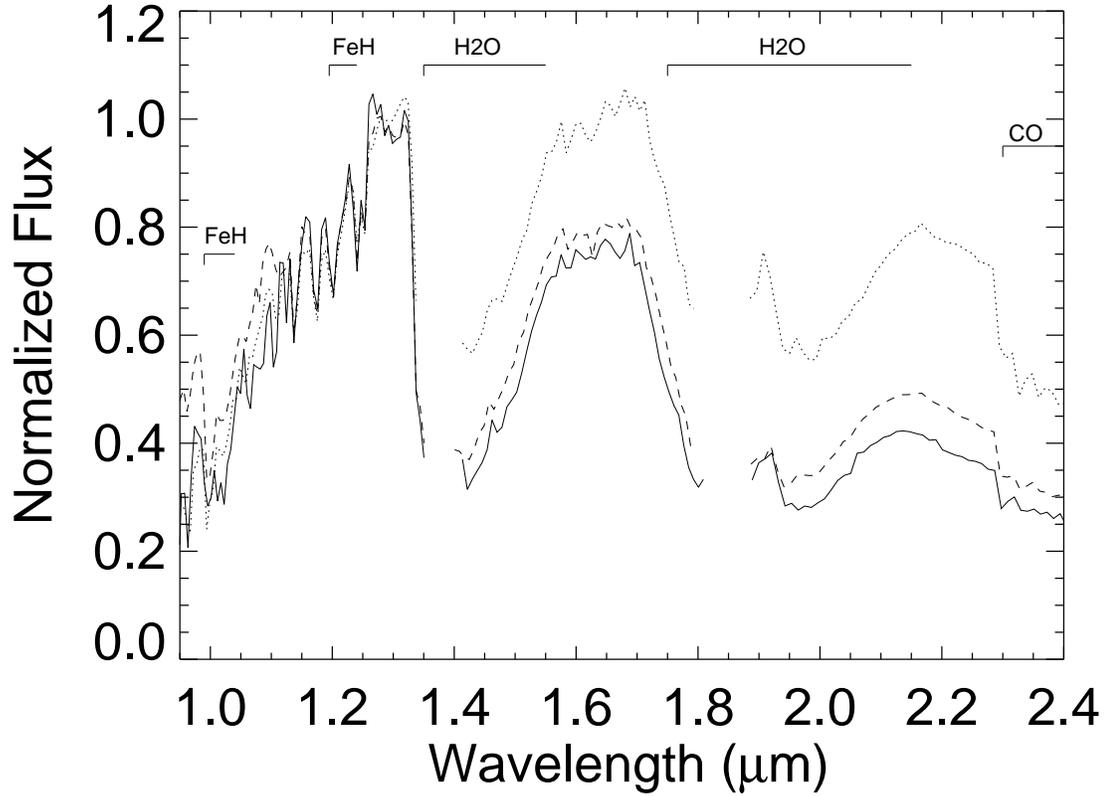}
\caption{Spectrum of blue ($J-K_s=1.28$) L dwarf 2M1711+4028 (solid
  line, this paper) overplotted with spectrum
  of relatively red ($J-K_s=2.05$) L4.5 dwarf 2MASS J22244381-0158521
  \citep[top dotted line,][]{kirkpatrick00}
  and relatively blue ($J-K_S=1.52$) L5 optical standard 2MASS
  J15074769-1627386 \citep[dashed line,][]{reid00}.  Both comparison spectra are from the IRTF Spectral Library.
  All spectra are normalized in 1.27-1.29~$\mu$m.  They have been rebinned to a
  lower resolution for clarity.
\label{fig:blue2}}
\end{figure}

The NIR colors of mid-late L dwarfs vary significantly within a single spectral
type.  For L5 dwarfs there is a spread of $\sim$0.7 magnitudes
in $J-K_s$ \citep{kirkpatrick08,cushing08}.  Although surface gravity
and metallicity play a role, comparison of atmospheric models to
actual spectra \citep[e.g.][]{knapp04, cushing08,
burgasser08} suggests that large variations in the NIR colors of L
dwarfs are primarily related to the properties of condensate clouds in their
atmospheres, with unusually red SEDs arising from thick clouds and
blue ones from thin or large-grained clouds.  
Common to the known peculiar blue L dwarfs is exaggerated H2O
absorption and diminished CO, as seen in the spectrum of 2M1711+4028.
As discussed in \S \ref{sect:secondary} the discrepancy between the
late-type H20 indices and the earlier type FeH and K1 indices, along
with its unusually blue NIR colors are indications that 2M1711+4028
falls into this category.  For comparison we overplot 2M1711+4028's spectrum with the
spectra of the very red L4.5 dwarf 2MASS J22244381-0158521 \citep{kirkpatrick00}, and the
relatively blue L5 optical standard 2MASS~J15074769-1627386 (see figure \ref{fig:blue2}).  All spectra
agree reasonably well in the J-band but diverge significantly at H and K, which may
indicate differing properties of condensate clouds in their atmospheres.
As a member of a wide binary system the surface gravity and
metallicity of 2M1711+4028 can be constrained from properties of the
primary, yielding an excellent laboratory for studying BD
atmospheres.  The estimated age of 1-5~Gyr for the G 203-50 primary
implies a relatively high surface gravity for the companion, possibly
contributing the its blue NIR colors.  However
surface gravity alone does not seem to be sufficient in explaining the
NIR colors of peculiar blue L
dwarfs \citep{burgasser08}.  Additionally, the
primary shows no signs of being metal poor, lending support to the
hypothesis that unusually blue NIR colors can be primarily attributed
to cloud properties. Higher resolution spectroscopy of the primary
and a more precise determination of the metallicity is required in
order to confirm this conclusion.

Another potential cause of unusually blue NIR colors is unresolved
binarity.  This may explain the slight onset of CH4 absorption at 2.2 $\mu$m in
the spectrum of 2M1711+4028.  At least one of the known peculiar blue L dwarfs, 2MASS
J08053189+4812330 \citep{burgasser07b} is thought to be an unresolved
binary with L4.5 and T5 components.  However, this system exhibits a
pronounced dip in the 1.6 $\mu$m CH4 feature due to the peaked shape of
the T dwarf's SED, whereas the spectrum of 2M1711+4028 is
relatively flat in that region.  Furthermore, the SED and colors of our
companion are very similar to that of another blue L4.5 dwarf, 2MASS
J11263991-5003550 (2M1126-5003 hereafter), discussed at length by
\citet{burgasser08}.  By constructing composite spectra using published L and T dwarf spectra from the SpeX
prism library, \citet{burgasser08} determined that no reasonable
composite spectrum could be found that matched that of 2M1126-5003 in
both the optical and NIR.  Without an optical spectrum for 2M1711+4028
we are limited in the conclusions we can draw, but its similarities to
2M1126-5003 may suggest that 2M1711+4028 is a single BD.
Our AO images support this conclusion, indicating that the BD is not a
near-equal mass binary with separation $> 0.18\arcsec$.

\subsection{Formation of G 203-50AB}\label{sect:formation}

\begin{figure}[here]
\epsscale{1}
\plotone{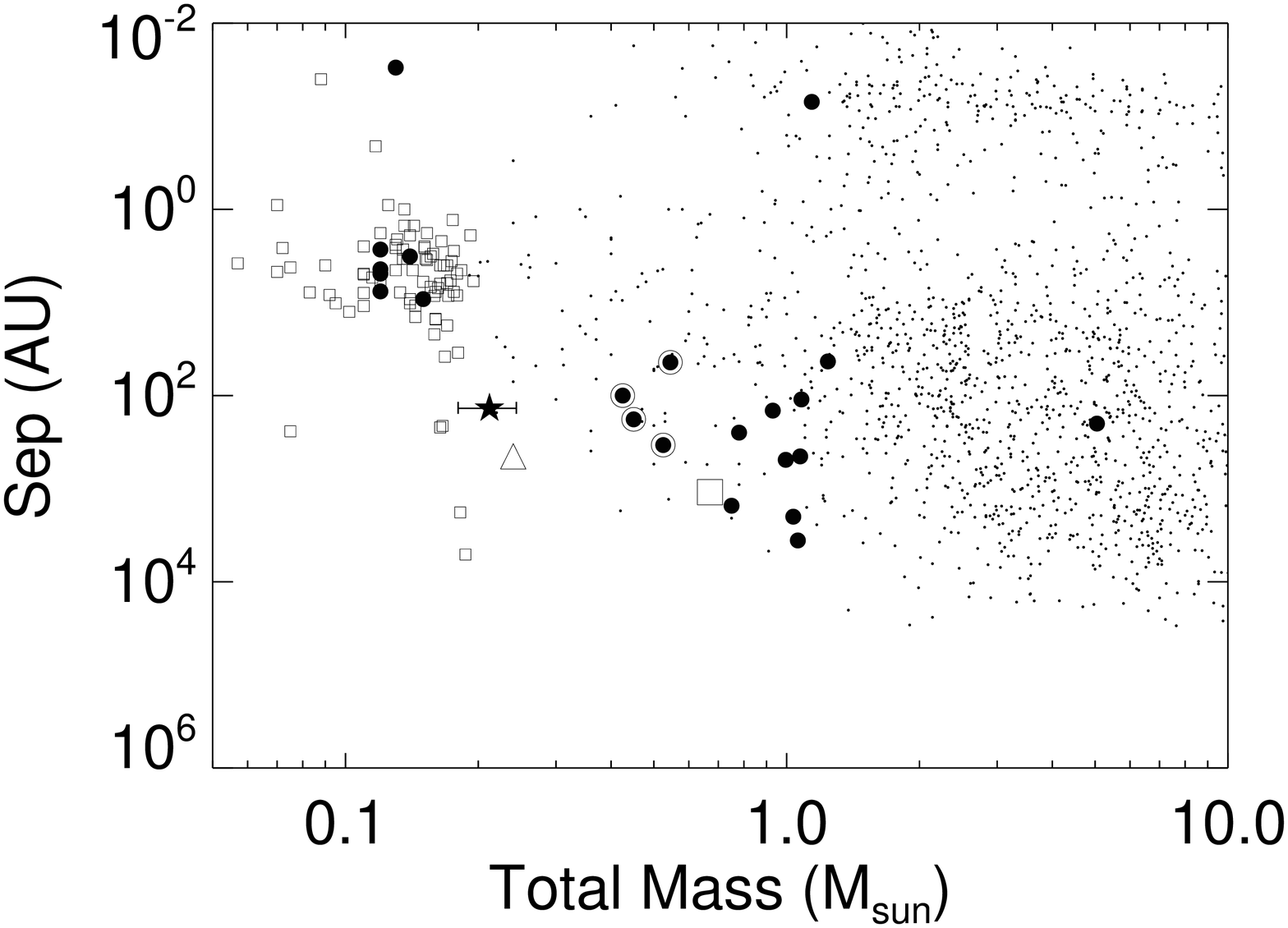}
\caption{Separation versus total system mass for known binary systems.
  Stellar binaries (dots) are from \citet{tokovinin97,FM92,reid97b,close90}; VLM
  binaries (open squares) are from the VLM binary archive, maintained by Nick
  Siegler (vlmbinaries.org); BD-stellar binaries (filled circles) are
  from \citet{reid01,met05}.  G~203-50AB (star symbol) falls in
  between the VLM and solar analogue regimes, and appears to be more
  loosely bound than  most systems of similar mass.  BD-stellar
  systems with M-dwarf primaries from \citet{reid01} have been
  circled.  The large triangle and square are BD-M dwarf systems from
  \citet{wilson01} and \citet{reid06}. 
\label{fig:sep}}
\end{figure}

With a total mass of $\sim$0.21~\msun~G~203-50AB is slightly more massive than the rare
wide VLM binaries, but much less massive than the solar analogues
around which BDs are routinely found at wide separations (see \S
\ref{sect:intro}, and figure \ref{fig:sep}).  It is
therefore of interest to consider how G~203-50AB may have
formed. Could the secondary have formed through
gravitational instability in a disk around the primary? Given the mass ratio
of $q$=0.45, that would imply a $M_{disk} >$ 0.45$M_*$, whereas typical
disks around low-mass stars contain a few percent of $M_*$ at $\sim$1 Myr \citep{scholz06}.
Since it is unlikely that the entire disk would end up in the companion, the
total disk mass, even in a conservative estimate, would have to be larger
than the primary's own mass to start with. Thus, we conclude that formation
of 2M1711+4028 in a protostellar disk around G~203-50 is implausible.

On the other hand, gravitational fragmentation of prestellar cores appears
to be capable of forming a wide variety of binary systems, depending on the
size, mass and angular momentum of the core \citep[e.g.][]{bate00}. However,
simulations usually have some difficulty producing binary stars with low
component masses and wide separations \citep[e.g.][]{bate03,goodwin04}. Some theoretical models invoke ejection from the parent cloud to
halt further accretion that would otherwise lead to higher masses. Given its
projected separation of 135$\pm$25~AU, the G~203-50AB binary has a binding
energy of 12.6$\pm3.8~\times$~$10^{-41}$~erg, placing it below the empirical
``minimum'' noted by \citet{close03} and \citet{burgasser07a}.  Thus, it is
unlikely to have survived such an
ejection.  We suggest that G~203-50AB most likely formed via fragmentation of an isolated core and
did not suffer strong dynamical interactions during the birth process or
subsequently.

\subsection{Search Sensitivity}\label{sect:sensitivity}
In order to assess the sensitivity of our search we simulated proper
motion distributions of M dwarfs distributed uniformly in a spherical volume out
to 25 pc, with tangential space velocities drawn from the distribution
of \cite{schmidt07}.  To be sensitive
to a particular M dwarf primary, its displacement between the 2MASS and
SDSS surveys had to be greater than the $3\sigma$ dispersion of all
other stars in the 4~${\rm deg^2}$ section of the sky in which it was
found.  For each such section of the sky, the time baseline between the
surveys was computed and used to determine the minimum proper motion
required for a detection.  We assumed that the population of M dwarfs
within 25 pc was uniformly distributed and assigned equal weight to
each 4 ${\rm deg^2}$ area of the sky.  Using the simulated proper motion
distributions the fraction of M dwarfs whose proper motions we could
have measured was determined for each section of the sky, giving an
average fraction of 0.58.  Adopting an M dwarf space
density ($9.0 < M_V < 13.0$ or roughly M0.5-M5.5) of $283.37
\times10^{-4}~{\rm pc^{-3}}$ \citep{reid02}, and given a search area of 7668
${\rm deg^2}$ of the sky, we should have been sensitive to approximately
$201\pm12$ early-mid M dwarfs within 25 pc.

Although in some cases we were able to recover binary systems with
separations $<$~$4\arcsec$, we conservatively put a lower limit of
$6\arcsec$ on our sensitivity, ensuring that components are well
separated.  The upper limit for separation is set by our search
radius, which extended to $120\arcsec$.  These limits correspond to
projected separations of 30-600~AU at 5~pc and 150-3000~AU at 25~pc.  Our sensitivity to
companions around each star was dictated by the mean 2MASS J-band
limiting magnitude (S/N=10) of $\sim16.5$, corresponding to a minimum mass of
$\sim0.06$~\msun~at 25~pc, assuming an age of 1-5 Gyr.  Other factors preventing us from finding
companions include poor astrometry due to saturation of the primary,
or low S/N of the secondary.  To estimate the number of
binaries missed we used SIMBAD and
DwarfArchives\footnote{http://DwarfArchives.org} to compile a list of
31 M-dwarfs and 38 BDs with previously measured proper motions large
enough to pass our cuts, and tested whether we
could measure the same proper motions using SDSS and 2MASS astrometry.
We found that $91\%$ of the time for M dwarfs, and $79\%$ of
the time for BDs  our measured proper motions agreed with the
previously measured ones, using the same criteria as our matching
algorithm described in \S\ref{sect:search}.  Therefore, we should have
been capable of identifying approximately $72\%$ of binaries
with sufficiently high proper motions.  Correspondingly we adjust our
sensitivity to $\sim145\pm9$ M dwarfs.  Adopting Poisson uncertainties on
a $1\sigma$ confidence interval for our single detection we roughly estimate that $0.7^{+1.5}_{-0.6}\%$
of early-mid M dwarfs have substellar companions with masses greater
than $\sim0.06$~\msun, at separations above $\sim120$~AU.

\section{Summary and Outlook}\label{sect:conclusions}

Above we have outlined our discovery of a wide, unusually blue, L5 companion to the nearby
M4.5 dwarf G 203-50.  Since BDs cool with time, it is not possible to infer their masses from observed luminosities.  In order to break this degeneracy, the age (or
an age indicator such as gravity) of the BD must be known.  Even so,
determining the mass of a BD requires accurate evolutionary models.
In order to constrain these models we must rely on a handful of BDs
for which independent age estimates can be obtained.  Our companion, 2M1711+4028 falls into this category, as its age can be constrained from the
age of the primary.  With
an angular separation of over 6\arcsec ~the components of G~203-50AB
are well separated, allowing the primary and secondary to be studied
independently.  At a distance of only $\sim21$~pc, the parallax can be
measured relatively easily, providing a more precise
determination of distance and luminosity.  Assuming an age
of 1-5 Gyr, 2M1711+4028 is older than most BDs with
independent age estimates (e.g. those found in star forming regions)
and can therefore provide an anchor point in a poorly constrained
mass-age regime.  Furthermore, as an unusually blue L dwarf in the NIR, 2M1711+4028
provides a unique opportunity for studying the relative importance of
gravity, metallicity and cloud properties in determining the NIR
colors of L dwarfs.   
  
With a total mass falling between the solar mass and VLM
regimes, G~203-50AB also has an important bearing on star formation theory.  Based
on the large mass ratio between the system components, we rule out
formation of the companion in the disk of the primary.  Instead we
suggest that this weakly bound binary formed via the fragmentation of
an isolated core, and did not suffer disruptive dynamical
interactions.  Statistically we put an upper limit of $2.2\%$ on the
wide companion fraction for BD companions with $m > 0.06$~\msun ~around
early-to-mid M dwarfs. 

In order to better constrain the properties of this unique system we
recommend that future observations of G~203-50AB include: a
parallax measurement to resolve uncertainties over the distance and
absolute magnitude of G~203-50; a high resolution spectrum of G~203-50
in order to determine the metallicity more precisely; an optical
spectrum for 2M1711+4028 to determine an optical spectral type; and time
series spectra to check for spectroscopic binarity.

\acknowledgments
We thank the anonymous referee for a prompt and
thoughtful review that greatly improved the quality of this manuscript.  We would like to thank the IRTF support staff, especially John Rayner
for walking us through SpeX observing procedures, and sharing with us
his valuable knowledge of the instrument.  We would
also like to thank Neil Reid for providing us with a spectrum for G
203-50.   This research has benefited
from the SIMBAD database,operated at CDS, Strasbourg, France; the M, L and T dwarf compendium
housed at DwarfArchives.org and maintained by Chris Gelino, Davy Kirkpatrick, and
Adam Burgasser;  The VLM Binaries Archive at VLMBinaries.org,
maintained by Nick Siegler; The SpeX Prism Spectral Libraries, maintained by Adam
Burgasser;  the IRTF Spectral Library maintained by John
Rayner; and the M dwarf standard spectra
made available by Kelle Cruz at
http://www.astro.caltech.edu/$\sim$kelle/M\_standards.  This publication makes use of data products
from 2MASS, which is a joint project of the University of
Massachusetts and the Infrared Processing and Analysis
Center/California Institute of Technology, funded by the National
Aeronautics and Space Administration and the National Science
Foundation.  This publication also makes use of data products from SDSS, which is funded by the Alfred P. Sloan Foundation, the Participating
Institutions, the National Science Foundation, the U.S. Department of
Energy, the National Aeronautics and Space Administration, the
Japanese Monbukagakusho, the Max Planck Society, and the Higher
Education Funding Council for England.  Observations were based in part on data collected at Subaru Telescope, which is operated by the National Astronomical Observatory of Japan. JR
is supported in part by an Ontario Graduate Scholarship.  DL is
supported in part through a postdoctoral fellowship from the Fonds Qu\'eb\'ecois de la Recherche
sur la Nature et les Technologies.  This work was supported in part
through grants to RJ and RD from the Natural Sciences and Engineering
Research Council (NSERC), Canada, and an Early Researcher Award from
the province of Ontario to RJ.

\clearpage

\begin{deluxetable}{lcc}
\tablewidth{0pt}
\tabletypesize{\small}
\tablecolumns{15}
\tablecaption{Physical properties of the components of G 203-50AB\label{tbl:comp}}
\tablehead{
\colhead{Quantity} & \colhead{A} & \colhead{b}
}
\startdata
Designation               &  G 203-50               &  2MASS J17114559+4028578 \\
$\mu_{\alpha}\cos{\delta}$ (mas~${\rm yr^{-1}}$)&  $229\pm43$      &  $256\pm36$ \\
$\mu_{\delta}$ (mas~${\rm yr^{-1}}$)       &  $61\pm49$        &  $92\pm43$ \\
2MASS $J$  (mag)                       &  $11.074\pm0.016$         &  $15.00\pm0.06$ \\
2MASS $M_J$   (mag)                  &  $9.34\pm0.44$ \tablenotemark{a}
& $13.5^{+0.7}_{-0.5}$\tablenotemark{b} \\
$J-K_s$ &0.80  & 1.28\tablenotemark{c} \\
$J-H$  & 0.51 & 0.79\tablenotemark{c} \\
$H-K_s$ & 0.29 & 0.49\tablenotemark{c} \\
$d$      &  $22.2\pm4.5$ &$20.0\pm6.3$ \\
Spectral type             &  M4.5\tablenotemark{d}  &  L5$^{+2}_{-1.5}$ \\
$T_{eff}$ (K)              & $3114\pm125$                   &  $1700^{+210}_{-250}$\\
Mass      (\msun)                & 0.146$\pm 0.031$     & $0.066^{+0.008}_{-0.015}$ \\
$H\alpha$~EW (\AA)           & $3.8\pm0.5$      & -- \\ 
TiO5\tablenotemark{d}    &    0.34 &--\\
CaOH\tablenotemark{d}    &    0.32 &--\\
CaH2\tablenotemark{d}    &    0.35 &--\\
CaH3\tablenotemark{d}    &    0.60 &--
\enddata
\tablenotetext{a}{Based on TiO5, CaOH, and CaH2 spectral indices \citep{mtcn7}}
\tablenotetext{b}{Based on average absolute magnitudes by spectral type, see \citet{dahn02}}
\tablenotetext{c}{synthetic colors computed from the spectrum of 2M1711+4028 using relative spectral response curves and zero-magnitude fluxes given in \citet{cohen03}}
\tablenotetext{d}{Spectral type and indices from \citet{mtcn7}}
\end{deluxetable}

\clearpage

\begin{deluxetable}{lc}
\tablewidth{0pt}
\tabletypesize{\small}
\tablecolumns{15}
\tablecaption{System properties of G 203-50AB\label{tbl:sys}}
\tablehead{
\colhead{Quantity} & \colhead{Value}
}
\startdata
Distance (pc)\tablenotemark{a} & $21.2\pm3.9$ \\
Angular separation ($\arcsec$) & $6.40\pm0.02$ \\
Physical Separation (AU)& $135\pm25$ \\
$\mu_{\alpha} cos\delta$ (mas~${\rm yr^{-1}}$)\tablenotemark{a} & $242\pm15$ \\
$\mu_{\delta}$(mas~${\rm yr^{-1}}$)\tablenotemark{a}& $77\pm17$ \\
Total mass (\msun) & $0.212\pm0.032$ \\
Mass ratio & $0.45\pm0.14$ \\
Binding energy (erg) & $12.6\pm3.8~\times10^{-41}$
\enddata
\tablenotetext{a}{mean quantities of the components} 
\end{deluxetable}

\clearpage

\begin{deluxetable}{lccc}
\tablewidth{0pt}
\tabletypesize{\small}
\tablecolumns{15}
\tablecaption{Spectral indices measured for 2M1711+2048\label{tbl:tab3}}
\tablehead{
\colhead{Index} &  \colhead{Value} & \colhead{Spectral Type} & \colhead{Reference}
}
\startdata
H2O 1.5~$\mu$m & 1.73  &  L8   &  1 \\
CH4 2.2~$\mu$m & 1.09  &  L7   & 1 \\
K1             & 0.33  &  L4.5 & 2\\
H2O$^{A}$      & 0.49  &  L7.5 & 3\\
H2O$^{B}$      & 0.60  &  L6   & 3\\
H2O A          & 0.45  &  L7   &  4  \\
H2O B          & 0.76  &  L5.5 &  4   \\
J-FeH          & 0.81  &  $<$L6   & 4\\
z-FeH          & 0.46  &  $<$L6   & 4 
\enddata
\tablerefs{(1)\citet{geballe02}; (2)\citet{tokunaga99};
  (3)\citet{reid01b}; (4)\citet{mclean03}} 
\end{deluxetable}

\clearpage

\clearpage

\clearpage

\clearpage

\clearpage

\end{document}